\documentclass[]{revtex4}
\usepackage[colorlinks,bookmarksopen,bookmarksnumbered,citecolor=magenta,urlcolor=red]{hyperref}
\usepackage{graphicx}
\usepackage{amsmath}
\usepackage{times}
\usepackage{natbib}
\usepackage{color} 
\begin{document}

\title{Surface Ocean Enstrophy, Kinetic Energy Fluxes and Spectra from Satellite Altimetry}

\author{Hemant Khatri$^1$, Jai Sukhatme$^{2,3}$, Abhishek Kumar$^4$ and Mahendra K. Verma$^4$}
\affiliation{1. Department of Mathematics, Imperial College London, London SW7 2AZ, United Kingdom. \\ 2. Centre for Atmospheric and Oceanic Sciences, Indian Institute of Science,
Bangalore, India. \\ 3. Divecha Centre for Climate Change, Indian Institute of Science, Bangalore, India.\\ 4. Department of Physics, Indian Institute of Technology Kanpur, Kanpur, India.}

\begin{abstract} Enstrophy, kinetic energy (KE) fluxes and spectra 
are estimated in different parts of the mid-latitudinal oceans via
altimetry data. 
To begin with, using geostrophic currents
derived from sea-surface height anomaly data provided by AVISO, we confirm the presence of 
a strong inverse flux of surface KE at scales larger than approximately 250 km. 
We then compute enstrophy fluxes to help develop a clearer picture of the underlying dynamics 
at smaller scales,
i.e., 250 km to 100 km. Here, we observe a robust enstrophy cascading regime, wherein the enstrophy 
shows a large 
forward flux and the KE spectra follow an approximate
$k^{-3.5}$ power-law. 
Given the rotational character of the flow, 
not only is this large scale inverse KE and smaller scale forward enstrophy transfer scenario consistent with expectations from 
idealized studies of three-dimensional rapidly-rotating and strongly-stratified
turbulence, it also agrees with detailed analyses of spectra and fluxes in the upper level midlatitude troposphere. 
Decomposing the currents into components with greater and less than 100 day variability (referred to as seasonal and eddy, 
respectively), 
we find that, in addition to the eddy-eddy contribution, the seasonal-eddy and seasonal-seasonal fluxes play a significant role 
in the inverse (forward) flux of KE (enstrophy) at scales 
larger (smaller) than about 250 km.
Taken together, we suspect, it is quite possible that,
from about 250 km to 100 km,
the altimeter is capturing the relatively steep portion of a surface oceanic counterpart of the upper tropospheric Nastrom-Gage spectrum.
%\vskip 0.2 truecm
%\begin{center}
%{\bf Submitted to Journal of Physical Oceanography.}
%\end{center}
\end{abstract}
\vskip 0.25truecm

\maketitle

\section{Introduction}
For the past decade, satellite altimetry data has been used to estimate the interscale transfer and spectral 
distribution of surface kinetic energy, henceforth abbreviated as KE, in the oceans. 
Focusing on mesoscales in mid-latitudinal regions with high eddy activity (for example, near the Gulf Stream, the Kuroshio or the Agulhas currents), 
the flux of KE is seen to be scale dependent. In particular, it has been noted that surface 
KE tends to be transferred upscale for scales larger than the local deformation radius and downscale for smaller scales
\citep{scott2005direct, tulloch2011scales, arbic2014geostrophic, arbic2012nonlinear}. 
Mesoscale wavenumber spectra, on the other hand, are somewhat more diverse with spectral indices ranging from $-\frac{5}{3}$ to $-3$ depending on the 
region in consideration
\citep{stammer1997global,le2008altimeter,xu2012effects}. In fact, recent work using in-situ observations suggests that the scaling changes with 
season, and is modulated by
the strength of eddy activity \citep{callies2015seasonality}.
Interpreting these results in terms of 
the dynamics captured by the altimeter and more
fundamentally the nature of the actual dynamics of the upper ocean
has been the subject of numerous recent investigations 
\citep[see for example the discussions in,][]{lapeyre2009vertical,ferrari2010distribution}. 

As baroclinic modes are intensified near the surface \citep{wunsch1997vertical,smith2001scales}, 
it has been suggested that 
altimetry data mostly represents 
the first baroclinic mode in the ocean \citep{stammer1997global}. Indeed, energy is expected to concentrate in the first baroclinic mode 
due to an inverse transfer 
among the vertical modes \citep{fu1980nonlinear}. 
Given this, at first sight, the observed inverse transfer of surface KE at large scales was surprising, as classical quasigeostrophic (QG) baroclinic 
turbulence anticipates a forward cascade in the baroclinic mode with energy flowing towards the deformation scale 
\citep{salmon1980baroclinic, hoyer1982closure}. 
However, a careful examination of the energy budget in numerical simulations reveals that, while KE goes to larger scales, 
the total energy in the first baroclinic mode does indeed flow downscale \citep{scott2007spectral}; a feature that is comforting in the context 
of traditional theory. In fact, along with the two-layer QG study of \citet{scott2007spectral}, 
inverse transfer of KE has also been documented in more comprehensive ocean models 
\citep{schlosser2007diagnosing,venaille2011baroclinic,arbic2012nonlinear,arbic2014geostrophic}. 

Noting the significance of surface buoyancy gradients (a fact missed in the aforementioned 
first baroclinic mode framework), it has been suggested that, surface QG (SQG) dynamics is a more appropriate framework 
for the oceans' surface \citep{lapeyre2006sqg}, and is reflected in the altimeter 
measurements \citep{lapeyre2009vertical}. Even though the variance of buoyancy is transferred downscale \citep{PHS,chaos}, 
surface KE actually flows upscale in
SQG dynamics \citep{smithetal2002,capet2008surface}, consistent with the flux calculations using altimetry data. Of course, 
in the QG limit, a combination of surface 
and interior modes is only natural. There are ongoing efforts to represent the variability of the surface ocean and interpret the
altimetry data in these terms 
\citep{lapeyre2009vertical,scott2012assessment,smith2012surface}.

Thus, much of the work using altimeter data has focussed on the inverse transfer of KE at relatively large scales. Here, we spend some time
on the larger scales, but 
mainly concentrate on slightly smaller scales that are still properly resolved by the data. Specifically, 
in addition to the KE flux, we also compute the spectral flux of enstrophy. In fact, this enstrophy flux sheds new light on the range 
of scales that span approximately 250 km to 100 km. We find that the enstrophy flux is strong and directed to small scales over this range,
and is accompanied by a KE spectrum that follows an approximate $k^{-3.5}$ power-law. 
This suggests that the rotational currents as derived from the altimeter
are in an enstrophy cascading regime from about 250 km to 100 km, and in an inverse KE transfer regime for scales greater than 
about 250 km. 
In terms of an eddy and slowly varying or seasonal decomposition (defined as smaller and larger than 100 day timescale variability, respectively), we observe that the 
seasonal-seasonal and seasonal-eddy fluxes play a significant role in the KE (enstrophy) flux for scales 
larger (smaller) that about 250 km.
Finally, we interpret these findings in the context of idealized 
studies of three-dimensional, rapidly rotating and stratified turbulence and also compare them with detailed analyses of 
midlatitude upper tropospheric 
KE spectra and fluxes.

\section{Data Analysis and Methodology}
\subsection{Data Description}
Gridded data of sea-surface height (SSH) anomalies (MADT delay time gridded data) from the AVISO project has been used in our analysis. 
%The SSH anomalies are computed with 
%reference to a long-term 21 year (1993-2013) mean. 
The data spanning 21 years (1993-2013) is available at a spatial resolution of 
$0.25^{\circ} \times 0.25^{\circ}$, thus scales smaller than approximately 50 km can not be resolved. 
We compute horizontal currents from the SSH anomaly data using geostrophic balance relations, and the 
latitudinal variation of the Coriolis parameter has been included in the computations.
 
\subsection{Geographical Locations}
\begin{figure}
\begin{center}
\includegraphics[width=9.5cm]{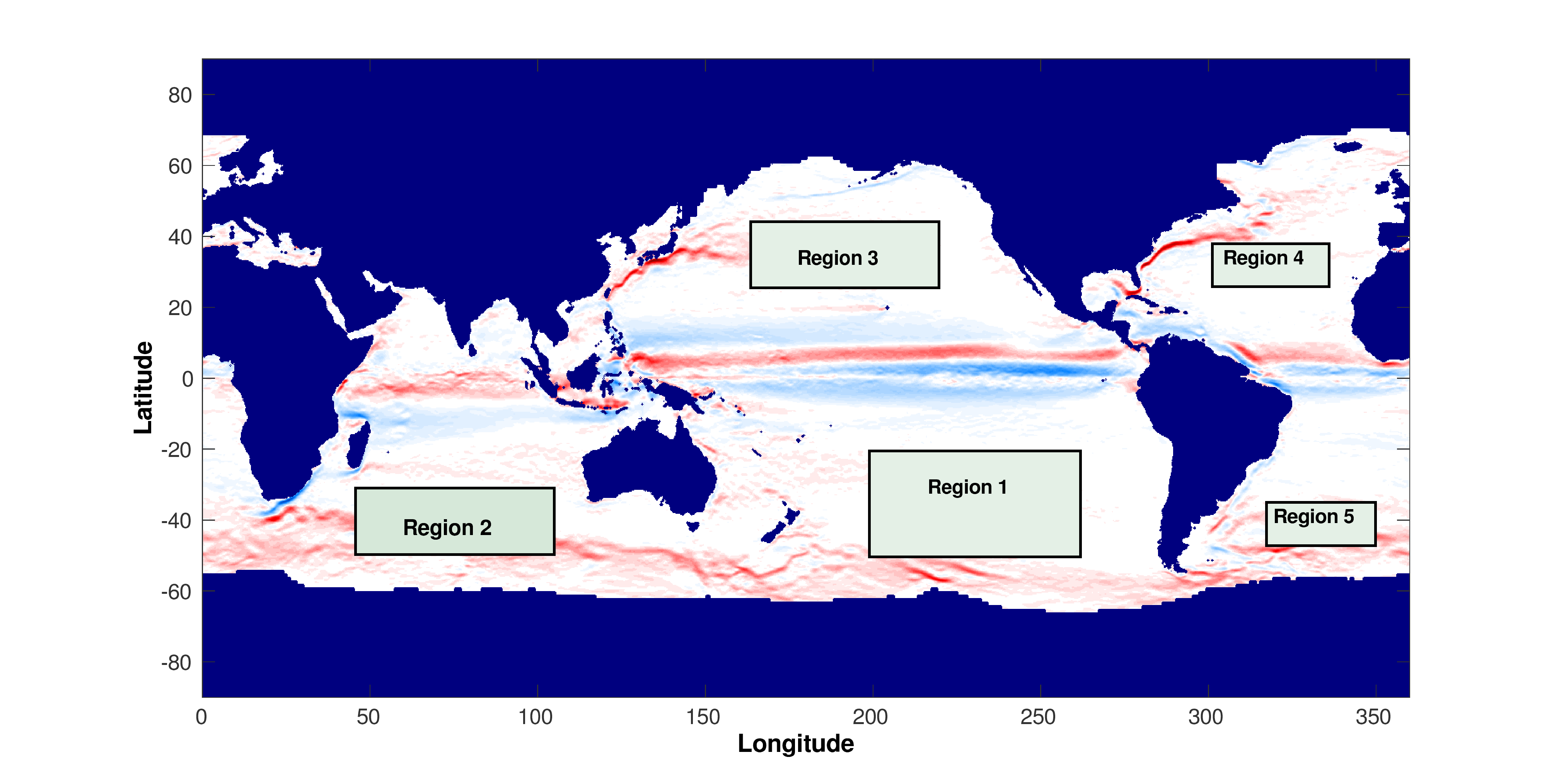}
\caption{Regions 1-5 chosen for the analysis. These are centered at $-36^{\circ}$, $232^{\circ}$ (grid size $256\times 128$);
$-38^{\circ}$, $78^{\circ}$ (grid size $256\times 64$); $33^{\circ}$, $187^{\circ}$ (grid size $256\times 64$);
$30^{\circ}$, $316^{\circ}$ (grid size $128\times 64$) and $-42^{\circ}$,  $336^{\circ}$ (grid size $128\times 64$), respectively.}
\label{fig:f1}
\end{center}
\end{figure}

The five geographical regions chosen for analysis are located far from the equator so that 
geostrophic balance is expected to be dominant. 
As is seen in Figure \ref{fig:f1}, these regions represent relatively uninterrupted stretches in the Northern and Southern parts of the Pacific, 
Atlantic, and the Southern Indian Ocean. In essence, we expect that this choice of domain minimizes boundary effects. Region 1 is the largest which is 
about 4500 km long and 3500 km wide. Other regions are comparatively smaller. 

\subsection{Computation of Spectra and Fluxes}
In this paper, we compute the KE spectrum and fluxes of KE and enstrophy. For this purpose, the data is 
represented using Fourier modes in both spatial directions, i.e., 
\begin{equation}
{{\bf U}}({\bf k}) = \int \!\!\! \int {\bf U}(x,y) \exp[-{\rm i}(k_x x + k_y y)] dxdy,
\end{equation}
where, ${\bf k}=(k_x, k_y)$ and ${\bf U}=(u,v)$ contains the zonal and meridional components of the velocity. 
The two-dimensional (2D) Fourier transform technique requires the data to have uniform 
grid spacing in both directions, so a linear interpolation scheme is employed to generate the velocity data on a rectangular grid 
(the original data is on equidistant latitudes and longitudes). 
In order to make the velocity field spatially periodic, the data is multiplied with a 2D bump function ($\exp[{-\frac{0.01}{1-x^2}-\frac{0.01}{1-y^2}+0.02}]$ where $(x,y)\in [-1,1]$) before performing a Fourier transform, this ensures that the velocity 
smoothly goes to zero at the boundaries.

In Fourier space, the KE equation is represented as \citep{scott2005direct},
\begin{equation}
\frac{\partial E(k)}{\partial t} = T(k) + F(k) - D(k),
\end{equation}
where $E(k)$ is the KE of a shell of wavenumber $k=\sqrt{k^2_{x}+k^2_{y}}$, $F(k)$ is the energy supply rate to the above shell by forcing, and $D(k)$ is the energy dissipation at the 
shell. $T(k)$ is the energy supply to this shell via nonlinear transfer. Note that, 
\begin{equation}
E(k) = \frac{1}{2} \!\!\sum_{k-1 < |k'| \le k} \!\!\!\!\!\!\!\!\!|{{\bf U}}({\bf k}')|^2.
\end{equation}
In a statistically steady state, $\partial E(k)/\partial t = 0$, and $E(k)$ is approximately constant in time. 
The energy supply rate due to non-linearity is balanced by 
$F(k)-D(k)$. A useful quantity called the KE flux, $\Pi(k)$, measures the energy passing through a wavenumber of radius $k$, and it is defined as, 
\begin{equation}
\Pi(k) = -\int_0^{k} T(k')dk'. 
\end{equation}
In 2D flows, another quantity of interest is the enstrophy ($\zeta = \int \! \frac{\omega^2}{2} d{\bf r}$), where $\omega$ is the relative vorticity. The corresponding enstrophy flux is denoted by $\zeta(k)$. 
We compute the KE and enstrophy fluxes using the formalism of \citet{dar2001energy} and \citet{verma2004statistical}, and the relevant formulae read, 
\begin{equation}
\Pi(k) = \sum_{k'> k}^{} \sum_{p \le k}^{} \delta_{\bf{k'},\bf{p}+\bf{q}}Im([\bf{k' \cdot U(q)}][\bf{U^{*}(k') \cdot U(p)}]), \label{eq:KE_flux}
\end{equation} 
 \begin{equation}
\zeta(k) = \sum_{k'> k}^{} \sum_{p\le k}^{} \delta_{\bf{k'},\bf{p}+\bf{q}}Im([\bf{k' \cdot U(q)}][\omega^{*}(\bf{k'}) \omega(\bf{p})]), \label{eq:w_flux}
\end{equation}
where $Im$ stands for the imaginary part of the argument and $\bf{U}^*$ is the complex conjugate. The expression (inside the summation operators) in equation \ref{eq:KE_flux} (\ref{eq:w_flux}) represents the energy (enstrophy) transfer in a triad (${\bf{k'}}={\bf{p}}+{\bf{q}}$) where $\bf{k'}$ mode receives energy (enstrophy) from modes $\bf{p}$ and $\bf{q}$. Then, the expression is integrated over all such possible triads satisfying the condition $k'> k$ and $p \le k$ (note that $-\int_0^k T(k')dk'$ = $\int_k^\infty T(k')dk'$).

\section{Results}

\begin{figure}
\begin{center}
\includegraphics[width=12cm]{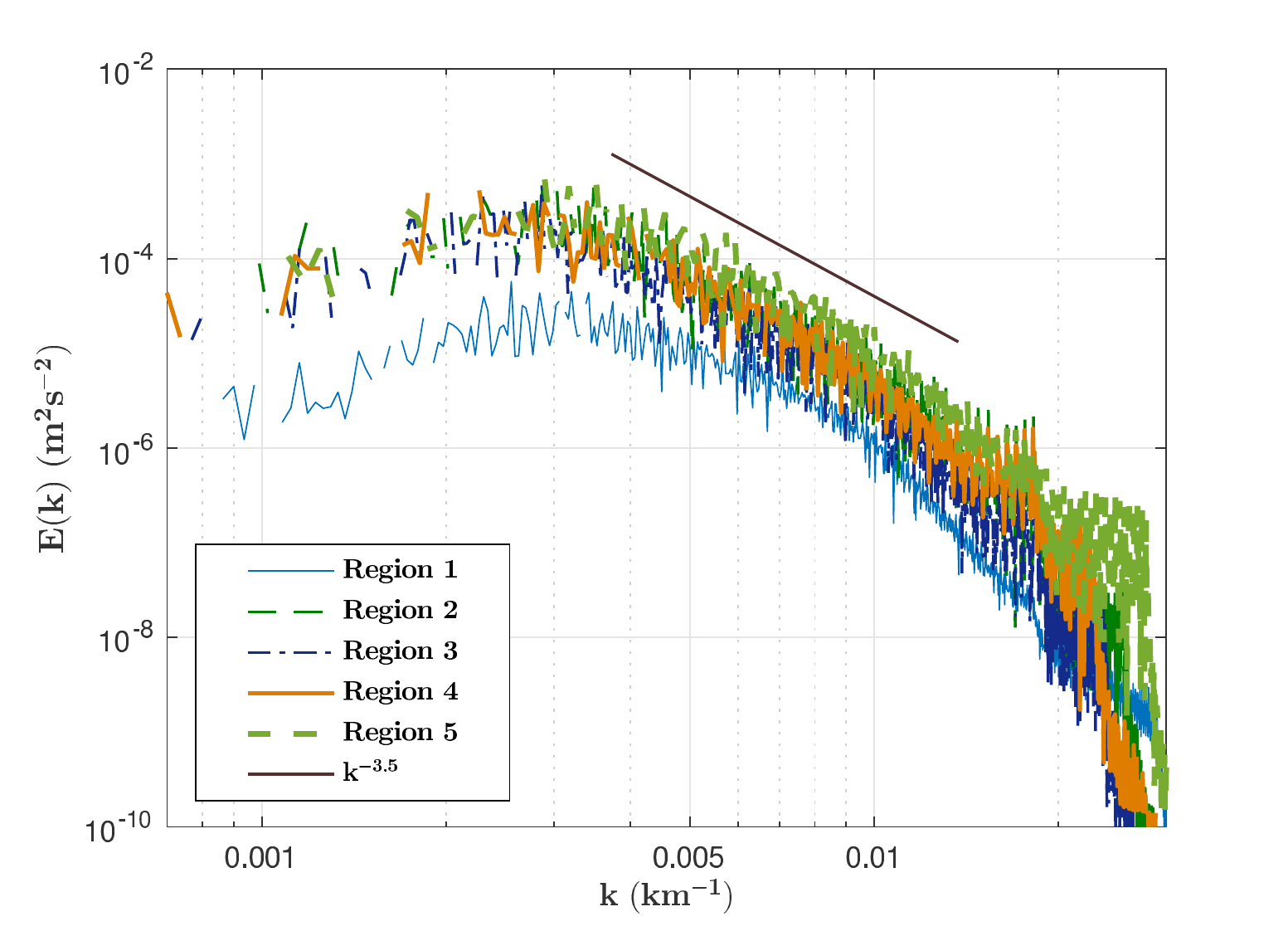}
\caption{Surface KE spectra for the geostrophic currents (averaged over 21 years of data). The slopes obtained between wavenumbers 0.004 and 0.01
(i.e., 250 km to 100 km) are:
$-2.85, -3.54, -3.48, -3.71$ and $-3.61$ for regions 1-5 respectively.}
\label{fig:f2}
\end{center}
\end{figure}

We begin by considering the spectra and fluxes associated with daily geostrophic currents.
Figure \ref{fig:f2} shows KE spectra of the geostrophic currents derived from SSH anomalies in all five regions. 
As seen, these currents show an approximate $k^{-3.5}$ scaling 
(the best fits range from $k^{-3.5}$ to $k^{-3.6}$) over a range of 250 to 100 km in all regions except Region 1 (the Southern Pacific), 
where the slope is somewhat shallower with a best fit of $k^{-2.9}$. 
Thus, the spectra we obtain for geostrophic currents are more in line with those reported by \cite{stammer1997global}, \cite{stam} (global extratropics), 
\cite{arbic2014geostrophic} (Agulhas region) and \cite{wang} \& \cite{callies} (near the Gulf Stream), but
differ from the shallower $-\frac{5}{3}$ like scaling observed by 
\cite{le2008altimeter} \citep[see also][]{xu2012effects}. 

\begin{figure}
\begin{center}
\includegraphics[width=12cm]{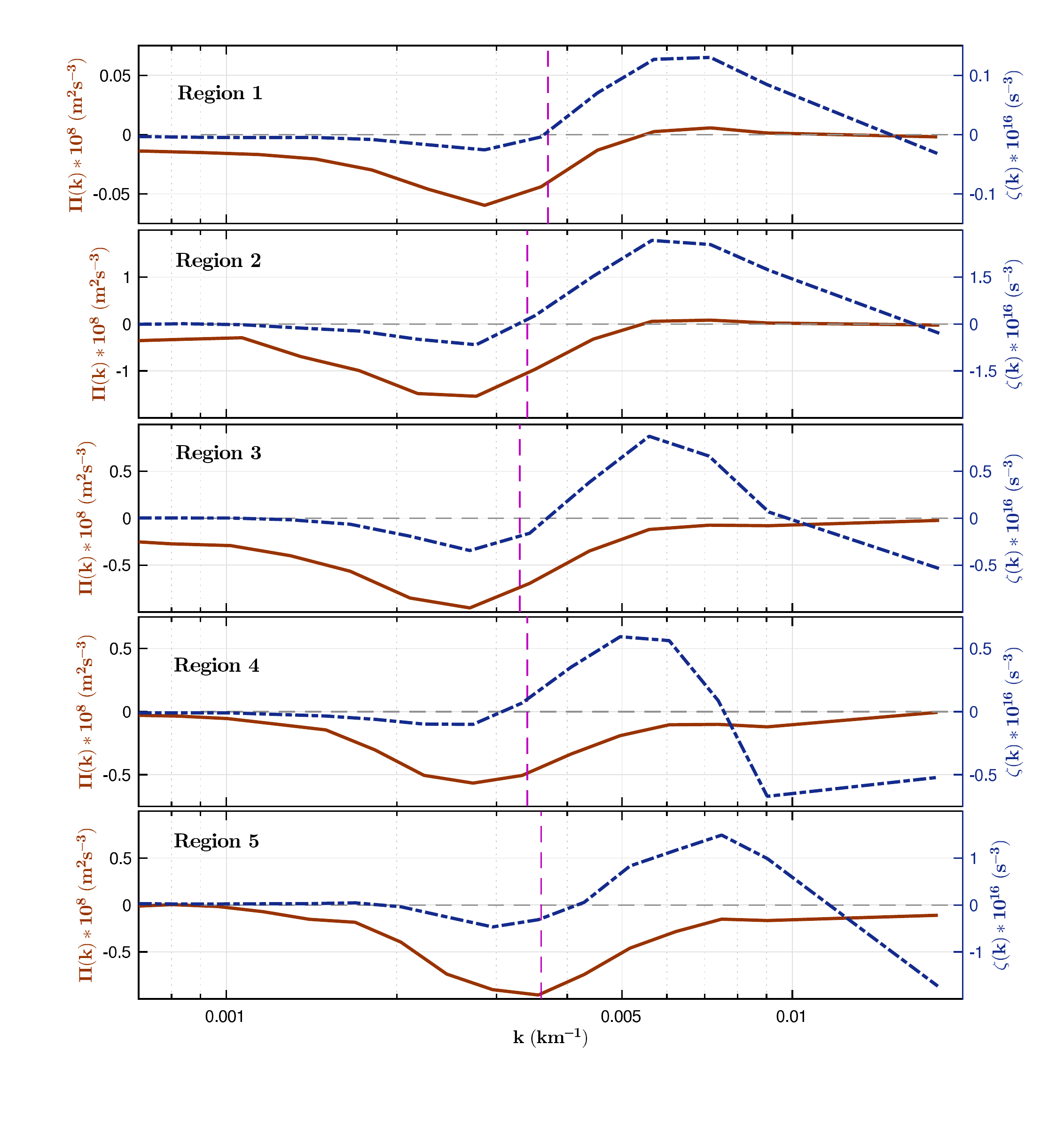}
\caption{KE (solid) and enstrophy (dash) fluxes for geostrophic currents (averaged over all 21 years) vs wavenumber ($k^2$ = $k^2_{x}$ + $k^2_{y}$).
Vertical dashed lines mark the transition scale ($\sqrt{\langle E \rangle/\langle \zeta \rangle}$)
in each region.}
\label{fig:f3}
\end{center}
\end{figure}

The KE and enstrophy fluxes are shown in Figure \ref{fig:f3}. 
Note that we have computed the flux using daily data and the results presented are an average over the entire 21 year period. 
The qualitative structure of the KE flux confirms the findings of \cite{scott2005direct} 
\cite[see also][]{scott2007spectral,tulloch2011scales,arbic2014geostrophic}. Specifically, 
we observe a robust inverse transfer of KE at large scales (i.e., greater than approximately 250 km). 
Some regions (1 and 2) show a very weak forward transfer of KE at small scales, while 
in the others (Region 3, 4 and 5), the flux continues to be negative (though very small in magnitude) even at small scales. 
In fact, in all the regions considered, the KE flux crosses zero or becomes very small by about 200 km.
Note that $2\pi$ times the climatological first baroclinic deformation scale in these five regions also
lies between 200 and 250 km \citep{chelton1998geographical}. Whether this is indicative of a KE injection scale due to linear instability as put forth by \cite{scott2005direct}, or more 
of a coincidence is not particularly clear.  
Indeed,
a mismatch between the deformation and 
zero-crossing scale can be seen in \cite{tulloch2011scales} and has also been pointed out by \cite{schlosser2007diagnosing}
in a comprehensive ocean model. 

Proceeding to the enstrophy, also shown in Figure \ref{fig:f3}, we see that it is characterized by a large forward flux at 
scales smaller than approximately 250 km. Further, the enstrophy flux does not show an inertial range, rather it increases with 
progressively smaller scales and peaks at approximately 150 km. Interestingly, 
we note that, in most of the regions, the 
scale $\sqrt{\langle E \rangle/\langle \zeta \rangle}$ (where $\langle \cdot \rangle$ denotes a domain average) \citep{danilov2000quasi} --- shown by the dashed vertical lines in Figure \ref{fig:f3} --- serves 
as a reasonable marker for the onset of the forward enstrophy flux regime.

\subsection{Eddy (subseasonal) and slowly varying (seasonal) fluxes}

\begin{figure}
\begin{center}
\includegraphics[width=12cm]{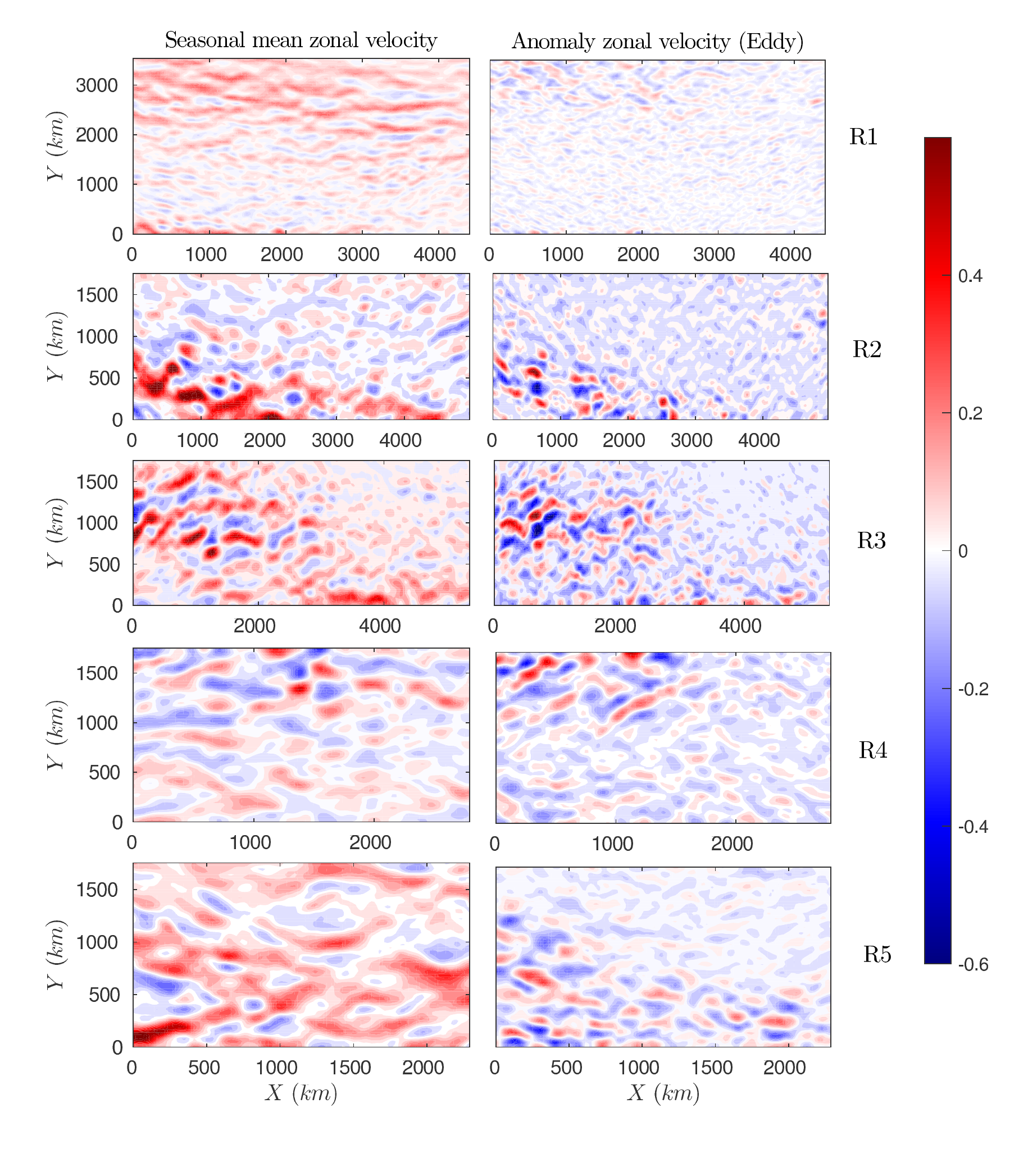}
\caption{Snapshot of the zonal seasonal and eddy fields in the summer (specifically on 16th June, 2005). R1-R5 correspond to Region1-Region5 of Figure \ref{fig:f1}.}
\label{fig:f4}
\end{center}
\end{figure}

\begin{figure}
\begin{center}
\includegraphics[width=12cm]{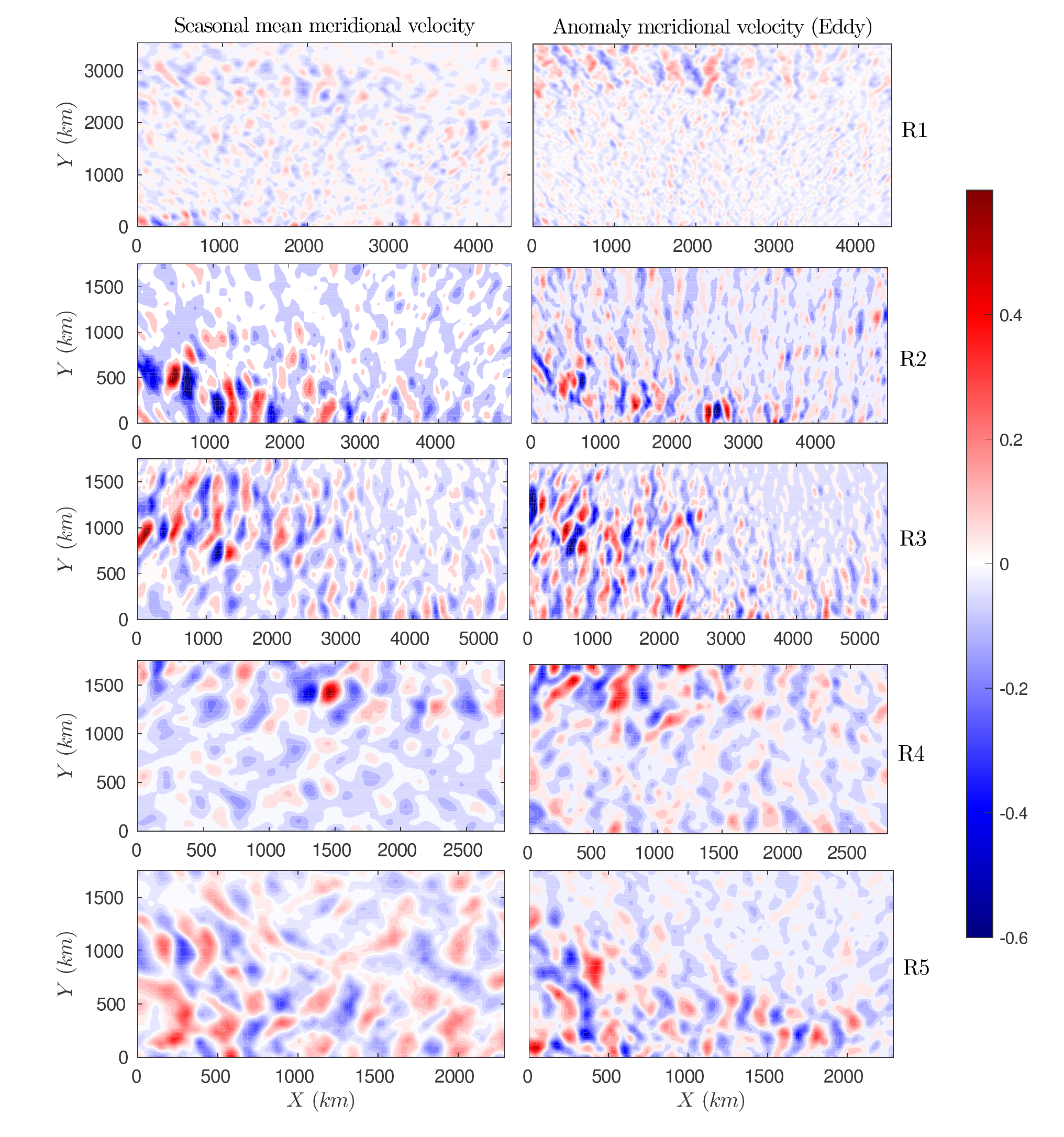}
\caption{Snapshot of the meridional seasonal and eddy fields in the summer (specifically on 16th June, 2005). R1-R5 correspond to Region1-Region5 of Figure \ref{fig:f1}.}
\label{fig:f5}
\end{center}
\end{figure}

An important difference between actual geophysical flows (the atmosphere and ocean) and idealized 3D rotating stratified turbulence
is the presence of nontrivial mean flows, and a hierarchy of prominent temporal scales. To get an idea of the KE and enstrophy 
flux contributions from the fast and slowly-varying components of the flow, following \cite{Shep}, we filter the 
derived geostrophic currents. In particular, at every grid point, we consider the daily 21 year long
time series and split this into two parts: one that contains variability of less than 100 days (referred to as the eddy or transient component) and 
the other with only larger than 100 day timescales (referred to as the slowly varying, or for brevity, as the seasonal component).
To get a feel for the physical character of the these decompositions, Figures \ref{fig:f4} and \ref{fig:f5} we show a snapshot of the zonal and 
meridional velocities (during summer) for all of the five regions. Quite clearly, the slowly varying or 
seasonal $u$ flow has a pronounced zonal structure, as compared to the more isotropic eddy component.
Similarly, the seasonal $v$ velocity has a larger scale and is oriented in a preferentially meridional direction as compared 
to its eddy component.
It is interesting to note that the meridional (seasonal and eddy) velocity is always comparable in strength to the zonal flow.

\begin{figure}
\begin{center}
\includegraphics[width=12cm]{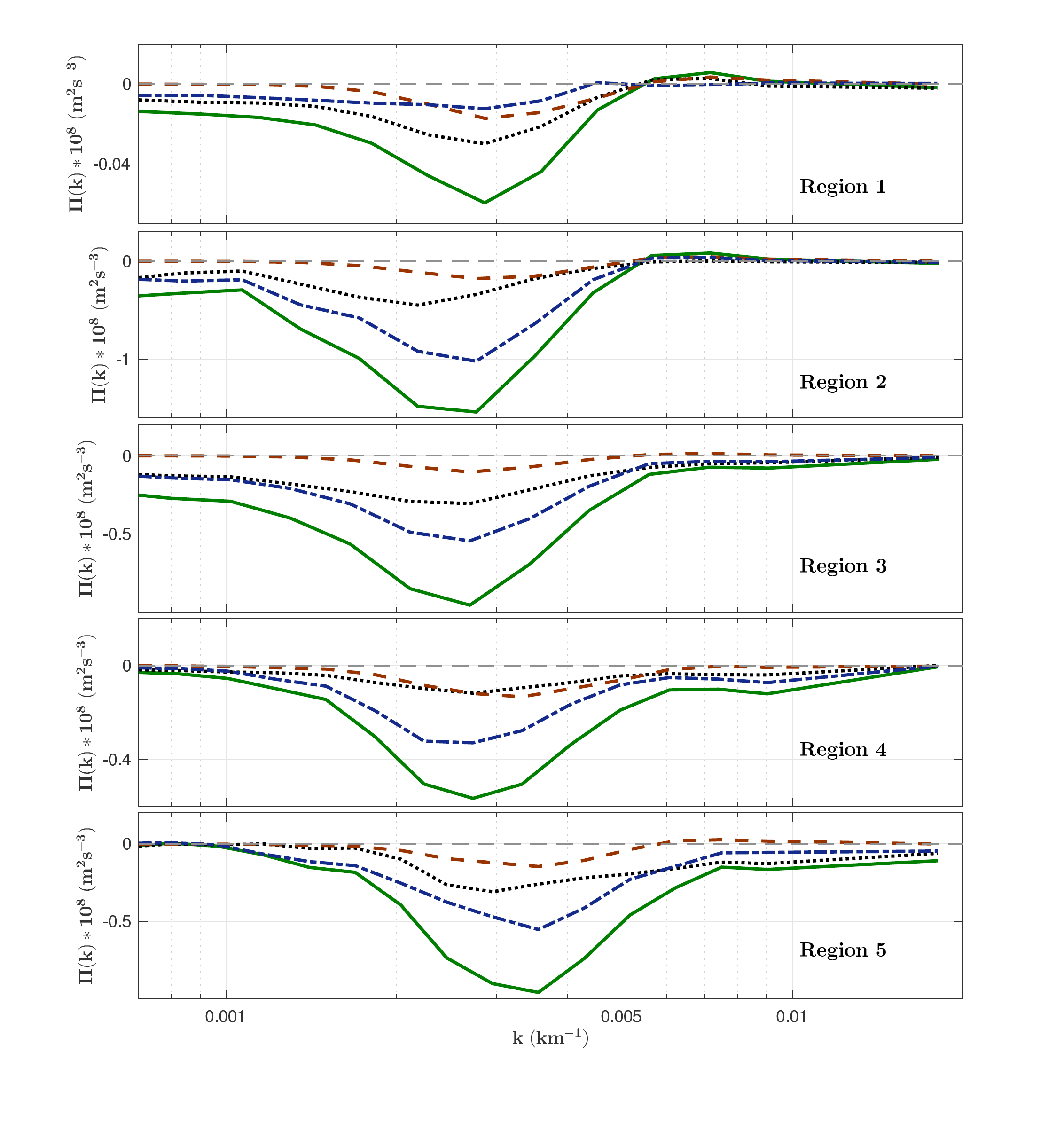}
\caption{Total (solid), eddy-eddy (dash), slowly varying-slowly varying (dot) and slowly varying-eddy (dash-dot) KE fluxes
for geostrophic currents (averaged over all 21 years) vs wavenumber ($k= \sqrt{k_x^2+k^2}$)
in all the five regions regions.}
\label{fig:f6}
\end{center}
\end{figure}

\begin{figure}
\begin{center}
\includegraphics[width=12cm]{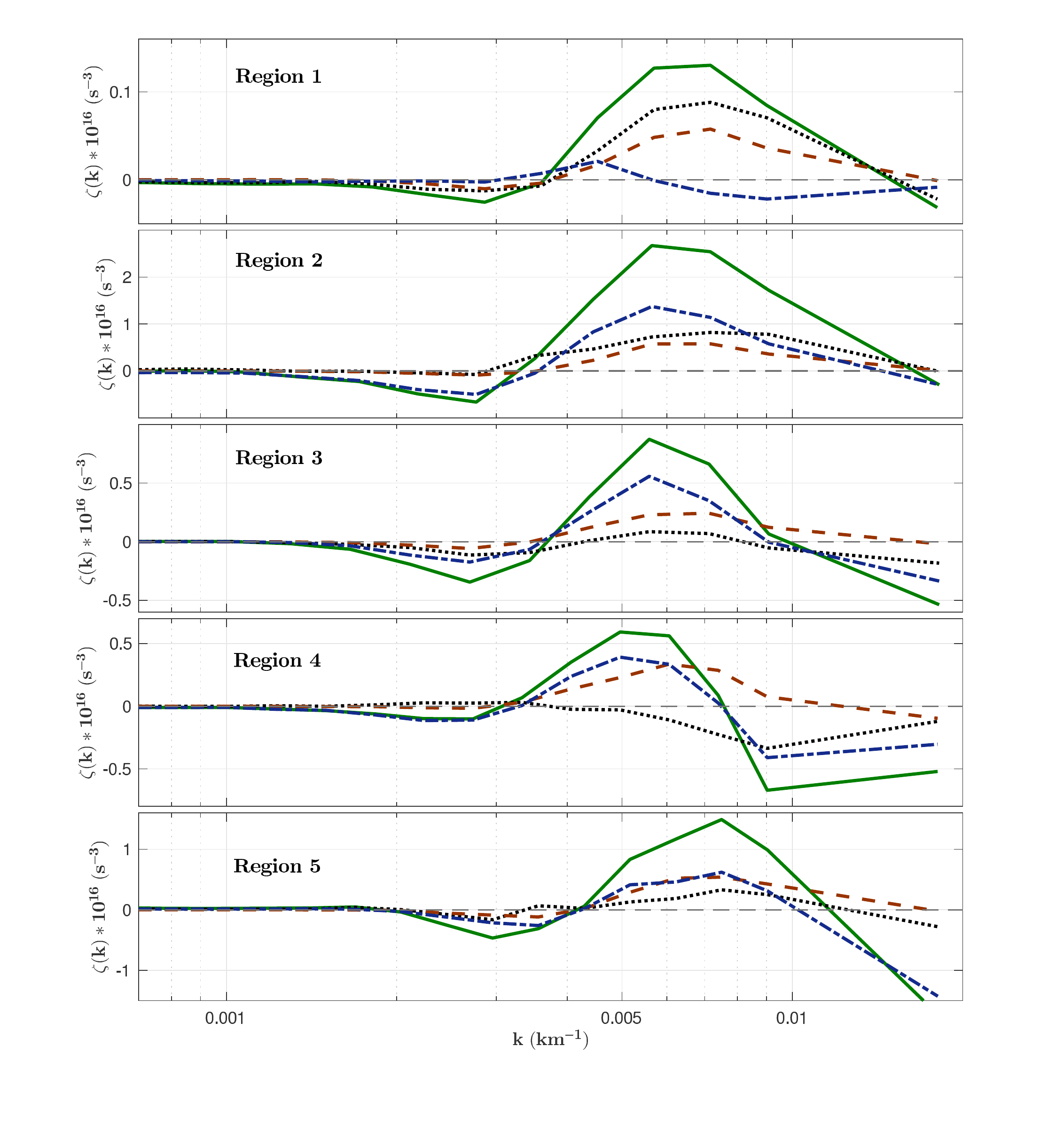}
\caption{Total (solid), eddy-eddy (dash), slowly varying-slowly varying (dot) and slowly varying-eddy (dash-dot) enstrophy fluxes for geostrophic currents (averaged over all 21 years) vs wavenumber ($k= \sqrt{k_x^2+k^2}$)
in all the five regions regions.}
\label{fig:f7}
\end{center}
\end{figure}

As with the original data (the total field), we compute the KE and enstrophy fluxes from the eddy field and the 
seasonal component. The seasonal-eddy fluxes are computed by subtracting the eddy-eddy and seasonal-seasonal contributions 
from the total flux.
Figures \ref{fig:f6} and \ref{fig:f7} show these four terms (total: solid curves, eddy-eddy: dashed curves, seasonal-seasonal: dotted curves and seasonal-eddy: dash-dot curves) 
for KE and enstrophy in the five regions 
considered, respectively. 
For the KE (Figure \ref{fig:f6}), we see that the total flux at large scales (i.e., greater than 250 km) has a strong contribution from the 
seasonal-eddy interactions. In fact, the eddy-eddy term 
is qualitatively of the correct 
form but quite small in magnitude. The seasonal-seasonal contribution is always upscale (except for a very small positive bump 
at small scales in Region 1), and thus, it too enhances the inverse transfer at large scales.
For the enstrophy (Figure \ref{fig:f7}), we see that the eddy-eddy term is reasonably strong, and along with the seasonal-eddy flux (in Regions 3,4 and 5), or the 
seasonal-seasonal flux (in Region 1), or both (Region 2) leads to the strong 
forward enstrophy cascading regime at small scales (i.e., below 
approximately 250 km).

\section{Interpretation and Conclusion}

By studying 21 years (1993-2013) of surface geostrophic currents derived from AVISO SSH anomalies in different midlatitudinal parts of the world's oceans
we find --- in agreement with previous studies --- that the spectral flux of rotational KE exhibits an inverse transfer at scales larger than about
250 km. Further, at smaller scales,
specifically, 250 km to 100 km, we find a strong forward flux of enstrophy accompanied by a KE spectrum that approximately follows a $k^{-3.5}$ power-law.
The KE flux at these small scales is very weak, and in a few of the regions considered, it is in the forward direction.
The transition from an inverse KE to a dominant forward enstrophy flux is roughly in agreement with a simple prescription based on the total enstrophy and KE in
the domain. On splitting the original data into high (eddy) and low (seasonal) frequencies, 
we observed that the seasonal-seasonal and seasonal-eddy 
fluxes play an important role in the 
inverse (forward) transfer of KE (enstrophy) at scales greater (smaller) than 250 km.
We now interpret these findings in the context of rapidly rotating, strongly
stratified three-dimensional (3D) turbulence as well as spectra and flux analyses from the midlatitude upper troposphere. 

Specifically, idealized 
3D rotating Boussinesq simulations suggest that rotational (or vortical) modes dominate the energy budget at large scales and exhibit
a robust inverse transfer of KE 
to larger 
scales, and a forward transfer of enstrophy to small scales \citep{bartello1995geostrophic,kitamura2006kh,sukhatme2008vortical,vall}. 
These transfers, akin to 2D and QG turbulence, are accompanied by KE spectra that follow $-\frac{5}{3}$ and $-3$ power-laws in the upscale KE and downscale 
enstrophy flux dominated regimes \citep{kraich,Charney}.
Given the rotational nature of the 
geostrophic currents, our observation of the upscale (downscale) KE (enstrophy) flux at scales greater (smaller) than 250 km is therefore in accord 
with the aforementioned expectations. Indeed, the $k^{-3.5}$ scaling is also close to the expected KE spectrum that characterizes the enstrophy flux dominated regime
\footnote{It should
be noted that the $-3$ exponent (even in incompressible 2D turbulence) is fairly delicate. As discussed in the review by \cite{boffetta2012two}, it is not
uncommon to observe power-laws for the KE spectrum that range from $-3$ to $-3.5$ in the enstrophy cascading regime.}.

With regard to the atmosphere, the forward enstrophy transfer regime of QG turbulence has been postulated to explain the $-3$ portion of 
midlatitude upper tropospheric KE spectrum \citep[the so-called Nastrom-Gage
spectrum,][]{NG}. Starting with \cite{Boer} and \cite{Shep}, re-analysis products at progressively finer resolutions have been analyzed
with a view towards seeing if the Nastrom-Gage spectrum is captured by the respective models \citep[see for example,][]{Strauss,koshyk,hamil}, 
and if so, what are the associated energy and enstrophy 
fluxes that go along with it \citep{lindborg-flux,Burgess}. In all, these studies demonstrate quite clearly that the $-3$ range of the Nastrom-Gage spectrum
(spanning approximately 3000-4000 km to 500 km in the upper troposphere) corresponds to the dominance of rotational modes, and a forward 
enstrophy cascading regime. Further, at scales greater than the $-3$ range (i.e., greater than approximately 4000 km), 
the upper troposphere supports an inverse rotational KE flux \citep{lindborg-flux,Burgess}.  

Regarding the small forward flux of rotational KE at scales smaller than approximately 250 km in a few regions, 
as pointed out by \cite{Boer} and \cite{Burgess}, this is 
likely due to the limited resolution of the data. For example, a forward KE flux in the rotational modes was observed in the coarse data used by
\cite{Boer}, but it vanishes in the more recent finer scale products analyzed in \cite{lindborg-flux} and \cite{Burgess} \citep[see,][for similar 
issues in incompressible 2D turbulence]{boff2}. In fact, 
much like the idealized 3D rotating stratified scenario \citep{bartello1995geostrophic,kitamura2006kh,sukhatme2008vortical}, the forward transfer of 
KE at small scales in atmospheric data (i.e., below approximately 500 km and accompanied by a shallower KE spectrum) is likely due to the divergent component of the flow \citep{lindborg-flux,Burgess}.

On decomposing the flow into eddy and seasonal components (less and greater than 100 day timescales, respectively), our results are somewhat
analogous to the upper troposphere. Specifically, in the atmosphere, the zonal mean-eddy (which translates to a 
stationary-eddy or seasonal-eddy decomposition) flux enhanced the inverse KE 
transfer to large scales \citep{Shep,Burgess}. We find the seasonal-eddy term to be important, but the 
seasonal-seasonal contribution to also be significant in the inverse transfer. In fact, in our decomposition (based on a timescale 
of 100 days), these terms 
dominate over the eddy-eddy contribution.
For enstrophy, in the upper troposphere, 
\cite{Shep} noted that the stationary-eddy fluxes are important (as we do here), 
but higher resolution data employed in \cite{Burgess} suggests that the eddy-eddy term is the dominant contributor to the 
forward enstrophy flux. 

Thus our findings using altimeter data, i.e., employing purely rotational
geostrophic currents, are in fair accordance with expectations from idealized simulations of rotating stratified flows as well as analyses of upper tropospheric
re-analysis data. 
The qualitative similarity in rotational KE fluxes, enstrophy fluxes, and KE spectra between the surface ocean currents and 
the near tropopause atmospheric flow is comforting as they both are examples of rapidly-rotating and strongly-stratified fluids.
In fact, in addition to an inverse rotational KE flux at large scales, we believe it is quite possible that the altimeter data is showing us
an enstrophy cascading, and relatively steep spectral KE scaling range of a surface oceanic counterpart to the atmospheric Nastrom-Gage spectrum. 
Quite naturally, it would be very interesting to obtain data at a finer scale, and see
if the ocean surface currents (rotational and divergent together) also exhibit a transition to shallower spectra --- like the upper tropospheric 
Nastrom-Gage spectrum --- with a change in scaling at a length scale 
smaller than 100 km.

\begin{acknowledgments}
We thank the AVISO project for making the SSH data  freely available (\url{http://www.aviso.altimetry.fr/en/home.html}). 
\end{acknowledgments}

\bibliography{ref}

\end{document}